\documentclass[prl,aps,twocolumn,groupaddress,longbibliography]{revtex4-1}
\usepackage{latexsym}
\usepackage{lineno}
\usepackage{hyperref}
\usepackage{url}
\usepackage{graphicx}
\usepackage{verbatim}
\usepackage{multirow}
\usepackage{amsmath}
\newcommand{\beq}{\begin{eqnarray}}
\newcommand{\eeq}{\end{eqnarray}}
\usepackage{mathrsfs}
\usepackage{float}
\usepackage[usenames, dvipsnames]{color}
\usepackage{mathtools}
\usepackage{slashed}
\usepackage{physics}	
\usepackage{graphicx}   
\usepackage{epstopdf}
\usepackage{subfigure}  
\usepackage{hyperref}   
\usepackage{bbold}
\usepackage{wasysym}
\usepackage{amssymb}
\usepackage{feynmp}
\usepackage[symbol]{footmisc}
\def \bs{\textbf}
\usepackage[latin1]{inputenc}
\usepackage{tikz}
\usetikzlibrary{decorations.pathmorphing}
\usetikzlibrary{shapes.misc}
\tikzset{cross/.style={cross out, draw=black, minimum size=8*(#1-\pgflinewidth), inner sep=0pt, outer sep=0pt},
cross/.default={1pt}}
\usetikzlibrary{patterns,math}
\newcommand{\RN}[1]{%
  \textup{\uppercase\expandafter{\romannumeral#1}}%
}
\begin{document}

\title{Dilute magnetic moments in an exactly solvable interacting host}
\author{Chandan Setty}
\thanks{email for correspondence: csetty@rice.edu}
 \affiliation{%
Department of Physics \& Astronomy,
Rice Center for Quantum Materials,
Rice University, Houston, Texas 77005,USA
 }%

\begin{abstract}
Despite concerted efforts, the problem of dilute local moments embedded in a correlated conduction electron host such as Nd$_{1-x}$Ce$_x$CuO$_2$ persists due to lack of analytically controllable models.  Here, we address the question: how do local moments couple to correlated but \textit{integrable} hosts? We describe the conduction electrons by the model of Hatsugai-Kohmoto (HK) which has undergone a recent resurgence arising from its exact solvability, existence of Luttinger surfaces, and connections to Sachdev-Ye-Kitaev (SYK) thermodynamics. We derive an exact low energy ``Kondo-HK" Hamiltonian 
and show the existence of additional spin-exchange coupling that is \textit{relevant} in the renormalization group (RG) sense. This term is ferromagnetic and does not vanish at low energies yielding an algebraic enhancement of the Kondo temperature.   
``Poor man's" scaling of couplings exhibits an exotic step-like RG flow between UV-IR fixed points attributed to severely restricted scattering phase space. This phenomenon is analogous to the flow of central charge in Zamolodchikov's diagonal resonance scattering in integrable quantum field theories.


\end{abstract}
\maketitle
\textit{Introduction:} When dilute local moments are placed in a metallic host, weakly interacting conduction electrons in the metal can screen the magnetic moment provided interactions between local moments are weak. The Kondo effect~\cite{Coleman2015} as it is termed, occurs due to the formation of a spin-singlet between the local moment and conduction electron spins below a characteristic crossover energy scale called the Kondo temperature given by $T_K \propto e^{-1/2 J N_0} $ ($J$ is the Kondo exchange coupling between the two spins and $N_0$ is the density of states). From a renormalization group (RG) standpoint, the coupling $J\equiv J(w)$ depends on the energy scale $w$ at which it is probed and corresponding exchange term in the effective Hamiltonian is marginal. Consequently, a perturbative ``poor man's" scaling of the coupling yields a $\beta(J)$ function of the form
\beq
\frac{d J}{d \ln w}\equiv \beta(J) = -2 J^2 N_0.
\label{beta}
\eeq  
 A key conclusion from the solution to Eq.~\ref{beta} at low energies is that $J(w)$ renormalizes to infinity (zero) if $J(w_0)>0$ ($J(w_0)<0$), where $w_0$ is the original conduction electron bandwidth.  Hence the formation of an antiferromagnetic spin-singlet quenches the local moment while no ferromagnetic coupling survives at low energy. \par
The aforementioned paradigm fails in the heavy fermion cuprate Nd$_{1-x}$Ce$_x$CuO$_2$~\cite{Czjzek1993, Zwicknagl1993} as electron correlations in the host render a breakdown of the quasiparticle concept. Hence the question of how local moments couple to strongly correlated hosts is highly non-trivial as one must either incorporate the conduction electron Hubbard or $t$-$J$ terms into the Anderson or Kondo models~\cite{Zwicknagl1993, Fulde1994, Fulde1994-2, Fulde1995, Fulde1995-2, Schork1996, Caron1996, Fazekas1996, Ishii1996, Tsvelik1998, Zevin1998, Vollhardt2000, Sakai1998, Zwicknagl2003, Rozenberg2008, Blawid1997, Igarashi1999,  Vollhardt2000, Kawakami2004, Fulde2007, Pruschke2007, Kawakami2011, Solyom2012, Senechal2018, Phillips2018}, rendering the problem notoriously difficult from the viewpoint of analytical tractability, or settle for reduced dimensionality of the host~\cite{Toner1992, Nagaosa1994, Li1995, PWP1996, Johannesson1996, Schiller1997, Fradkin1990}. \par In this paper, we pursue a simplification by raising the following alternate question: how do local moments couple to metallic hosts that are correlated but \textit{integrable}, and what is the fate of these couplings at low energy? A resolution of this question would be a significant step toward uncovering the thermodynamics of Kondo lattices in Nd$_{1-x}$Ce$_x$CuO$_2$ and its analogues. As the Hubbard model, minimal model applicable to the Cuprate normal state, is unsolved in two or more dimensions, we instead describe the conduction electrons by the Hatsugai-Kohmoto (HK) model~\cite{HK1992} (the acronym HK is not to be confused with the Hubbard-Kondo model routinely used in literature). 
The HK is a minimal model for a doped Mott insulator that is integrable -- there exist extensively many conserved quantities, one for each crystal momentum -- leading to exact solvability. The model has undergone a revival due to existence of Luttinger surfaces (LSs), contours of Green function zeros in momentum space. It was recently demonstrated that pairing instabilities on a LS and their corresponding fluctuation thermodynamics resemble that of the $O(1)$ Sachdev-Ye-Kitaev free energy~\cite{Setty2020, Setty2021}. Following these results, Ref.~\cite{Phillips2020} derived a Cooper instability by adding an infinitesimal attractive interaction to the HK model whereas Ref.~\cite{Yang2021} reproduced the Cuprate Fermi arcs and pseudo gap. The HK interaction term has also been used recently to drive topological transitions~\cite{Wang2021} and assess stability conditions for LSs to perturbations~\cite{Phillips2021}. In real space, the HK Hamiltonian comprises a quadratic nearest neighbor hopping term and a quartic interaction term that is non-zero only for electrons with a net zero center-of-mass. In momentum space, such a Hamiltonian acquires a particularly simple form
\beq
H_{HK} = \sum_{\bs k \sigma} \xi_{\bs k} c_{\bs k \sigma}^{\dagger} c_{\bs k \sigma} + \tilde{U} \sum_{\bs k} n_{\bs k \uparrow} n_{\bs k \downarrow} \label{HK},
\eeq
where $\xi_\bs k$ is the band dispersion relative to the Fermi energy $\epsilon_F$, $c_{\bs k \sigma}$ destroys an electron at momentum $\bs k$, spin $\sigma$ and follows fermion anti-commutation relations,  $n_{\bs k \sigma}$ is the number operator, and $\tilde{U}$ is the HK four fermion interaction which, unlike the Hubbard model, is local in momentum instead of position. A notable feature of Eq.~\ref{HK} is that kinetic and interaction terms commute and the particle number is conserved for each $\bs k$ rendering exact solvability of the model. The ground state can be written in the particle number representation as $|\psi_G, \{\sigma_j\}\rangle = \prod_{\bs k \in \Omega_1}c_{\bs k \sigma_j}^{\dagger}\prod_{\bs k \in \Omega_2} c_{\bs k \uparrow}^{\dagger}c_{\bs k \downarrow}^{\dagger}| 0 \rangle $ where $\Omega_i$ is the region in momentum space with occupation number $i$. The model exhibits a Mott gap (correlated metallic behavior) for $\tilde{U}>W$ ($\tilde{U}<W$) where $W$ is the non-interacting bandwidth.  In addition, the ground state has a large degeneracy resulting from all possible spin configurations $\{\sigma_j\}$ in the singly occupied regions of the Brillouin zone. The exact Green function is 
\beq
G_{\sigma}(\bs k, i\omega_n) = \frac{1- \langle n_{\bs k \bar{\sigma}}\rangle}{i\omega_n-\xi_{\bs k}} + \frac{\langle n_{\bs k \bar{\sigma}} \rangle}{i\omega_n-(\xi_{\bs k}+\tilde{U})},\label{GF}
\eeq
where $\langle n_{\bs k \bar{\sigma}}\rangle$ is the average occupation of state $\bs k$ with opposite spin $\bar{\sigma}$.  An examination of Eq.~\ref{GF} reveals a crucial simplification of the HK model -- $G_{\sigma}(\bs k, i\omega_n)$ is additive in doubly occupied and unoccupied sectors of the theory with no entanglement between them. This is a consequence of the ground state being an eigenstate of the number operator which leads to exact cancellation of the mixing terms.  A physical implication of this cancellation is the separation of Mott gap formation for $\tilde{U}>W$ and dynamical spectral weight transfer, two important ingredients of Mottness~\cite{Phillips2010}. While the two are inextricably intertwined in the Hubbard model, the HK model harbors only a Mott gap with no dynamical spectral weight transfer thus making it considerably more tractable. Nevertheless, the lack of a quasiparticle picture due to `doublon-holon' excitations, the presence of a Mott spectral gap, divergence of the self-energy leading to LSs, and violation of the Luttinger sum rule make the ground state of the HK model irreconcilable with a Fermi liquid for any $\tilde{U}\neq0$~\cite{HK1992, Phillips2020}. \par
Our results below demonstrate that an inclusion of HK interaction term into the Anderson impurity model 
leads to fundamental differences with the Kondo effect in a Fermi liquid.  Applying  the Schrieffer-Wolf transformation~\cite{SWTransform-1966} to the Anderson-HK model we derive a low energy ``Kondo-HK" model which -- in addition to the \textit{marginal} Kondo spin exchange ($J$) -- acquires a \textit{relevant} spin exchange term ($D$) as well as a six-fermion term ($E$) formed by coupling the electron density to spins.  Unlike the conventional spin exchange coupling which is antiferromagnetic ($J>0$), $D$ is always ferromagnetic ($D<0$) but does \textit{not} vanish at low energy. With only the $D$ term present, the Kondo temperature is enhanced by interactions yielding a power law dependence on the coupling  ($T_K \sim D$) as opposed to an exponential. When $J$ and $E$ terms are turned on, only the coupling between $J$ and $D$ survives among the mixed terms and the Kondo-log in $J$ is cut-off at low energies. Furthermore, the existence of extensively many conservation laws for the number operator forces exotic `step-like' RG flows analogous to diagonal $S$-matrix resonance scattering previously studied in well known integrable field theories in high energy physics~\cite{Zamolodchikov1991, Zamolodchikov1991-2, Zamolodchikov1991-3, Zamolodchikov1991-4, Zamolodchikov1991-5, Martins1992, Ravanini1993, Castro2000, Leclair2003, Castro2004, Takacs2015, Pimenta2017, Lee2020}. Our calculations present novel insights into an age-old but persistent question of how local moments couple to correlated hosts where a quasiparticle picture is absent, and into the nature of thermodynamics and transport in dirty correlated systems~\cite{PJH2009-RMP}\par 
\textit{Model:} We begin by including the HK interaction term in Eq.~\ref{HK} into the Anderson impurity model. The total Hamiltonian for the Anderson-HK model is given by
\beq \nonumber
H &=& H_0 + H_v \\ \nonumber
H_0 &=& \sum_{\bs k \sigma} \xi_{\bs k} c_{\bs k \sigma}^{\dagger} c_{\bs k \sigma} + \tilde{U} \sum_{\bs k} n_{\bs k \uparrow} n_{\bs k \downarrow} \\ \nonumber
&+& \sum_{\sigma} \epsilon_d d_{\sigma}^{\dagger} d_{\sigma}  + U n_{d\uparrow} n_{d\downarrow}\\ 
H_v &=& \sum_{\bs k \sigma} V_{\bs{k}} \left( c_{\bs k \sigma}^{\dagger} d_{\sigma} + d_{\sigma}^{\dagger} c_{\bs k \sigma}\right)\label{Model},
\eeq
where the local moment with spin $\sigma$ at the origin is created by the operator $d_{\sigma}^{\dagger}$ which follows fermion anti-commutation relations, and $n_{d\sigma}$ is the corresponding number operator.  We have defined the bare conduction electron dispersion $\xi_{\bs k} = \epsilon_{\bs k} -\epsilon_F$ and local moment energy $\epsilon_d$, measured with respect to $\epsilon_F$. $U$ is the onsite local moment repulsion and $V_\bs{k}\equiv V$ is the hybridization between the conduction electrons and local moments which is chosen to be a constant. 
Throughout the manuscript, we choose $\tilde{U} < W$ so that the conduction electrons are in the correlated metal phase.  For a Kondo-type interaction between the local moment and conduction electrons, we require a well-defined magnetic moment to exist. To his end, we choose to solve the problem in limit where $\tilde{U}, V \ll U$ so that the condition for local moment formation is not drastically altered from free electron case. This is ensured because the condition $\tilde{U}<W$ only weakly affects the density of states at the local moment energy ($\epsilon_d < \epsilon_F$), and local moment broadening still satisfies $\Delta \ll U$.  Despite the above hierarchy of scales and small $\tilde{U}$, the conduction electrons cannot be described by a Fermi liquid as long as $\tilde{U} \neq 0$.     
We now derive the Kondo-HK model from the Anderson-HK model Eq.~\ref{Model} by performing the Schrieffer-Wolf transformation on $H$ to eliminate double occupancy. The effective low-energy Kondo-HK Hamiltonian is given as $H_K = e^S H e^{-S}$ where the operator $S$ is obtained by imposing the condition that linear in $V$ terms vanish in $H_K$. This condition is given as $H_v + [S, H_0] =0$, 
so that the Kondo-HK Hamiltonian  becomes $H_K = H_0 + \frac{1}{2} [S, H_v] + \mathscr{O}(V^3)$ to quadratic order in $V$. To find the operator $S$, we choose the \textit{ansatz}
\beq \nonumber
S &=& V \sum_{\bs k \sigma} (A'_{\bs k} - B'_{\bs k} n_{d\bar{\sigma}} + D'_{\bs k} n_{\bs k \bar{\sigma}} + E'_{\bs k} n_{\bs k\bar{\sigma}} n_{d\bar{\sigma}}) \\ 
&&\left( c_{\bs k \sigma}^{\dagger} d_{\sigma} - d_{\sigma} c_{\bs k \sigma}\right)
\eeq
and substitute it into the condition $H_v + [S, H_0] =0$.  Solving for the primed coefficients we have $A'_{\bs k} = (\xi_{\bs k} - \epsilon_d )^{-1}$, $B'_{\bs k} = - U (\epsilon_d - \xi_{\bs k})^{-1} (U + \epsilon_d - \xi_{\bs k})^-1$, $D'_{\bs k} = \tilde{U} (\epsilon_d - \xi_{\bs k})^{-1} (\tilde{U} -\epsilon_d + \xi_{\bs k})^{-1}$ and $E'_{\bs k} = \frac{1}{U + \epsilon_d - \xi_{\bs k}} + \frac{1}{-\tilde{U} + \epsilon_d - \xi_{\bs k}} +\frac{1}{\xi_{\bs k} -\epsilon_d} + \frac{1}{-U + \tilde{U} -\epsilon_d + \xi_{\bs k}}$.
Setting $\tilde{U}$ to zero, we see that $D_{\bs k}', E_{\bs k}'$ vanish reducing the expression for $S$ to that of local moments in a Fermi liquid. With the expression for $S$ at hand, we can now evaluate the commutator $[S, H_v]$. Keeping only the spin-flip terms between the local moment and conduction electrons,  the effective low energy Kondo-HK Hamiltonian takes the form
\beq \nonumber
H_K&=& \sum_{\bs k \sigma} \xi_{\bs k} c_{\bs k \sigma}^{\dagger} c_{\bs k \sigma} + \tilde{U} \sum_{\bs k} n_{\bs k \uparrow} n_{\bs k \downarrow} \\  \nonumber
&+& \sum_{\bs k \bs k' } \left(J(\bs k, \bs k') + \delta_{\bs k, \bs k'} D_{\bs k}\right) c_{\bs k \mu}^{\dagger} \vec{\sigma}_{\mu \nu}c_{\bs k' \nu}\cdot \mathbf{S}_d \\ 
&+& \sum_{\bs k \bs k'} E_{\bs k} \left(n_{\bs k \uparrow} S^+_c(\bs k')S^-_d+n_{\bs k \downarrow} S^-_c(\bs k')S^+_d\right).
\label{EffectiveKondo}
\eeq
Here $\vec{\sigma}$ is the vector of $c_{\bs k \sigma}$ electron Pauli matrices, $\mathbf{S}_{c}(\bs k)$ and $\mathbf{S}_d$ are the spin operators for the conduction electrons and local moment respectively, $S^{\pm}_c(\bs k'), S^{\pm}_d$ are the corresponding components of the ladder operator,  $D_{\bs k} \equiv V^2 D_{\bs k}'$, $E_{\bs k} \equiv V^2 E_{\bs k}'$  and $J(\bs k, \bs k') \equiv - \left[\frac{V^2 U}{(\xi_{\bs k}- \epsilon_d)(\xi_{\bs k} -\epsilon_d -U)} + \bs k \rightarrow \bs k'\right]$. The first two terms in Eq.~\ref{EffectiveKondo} form the HK Hamiltonian, Eq.~\ref{HK}. The $J(\bs k, \bs k') $ term is the familiar spin-exchange Kondo term in a Fermi liquid. The terms proportional to $D_{\bs k}$ and $E_{\bs k}$ are entirely due to the HK interaction term and are absent in the Fermi liquid Kondo effect. Like the $J(\bs k, \bs k') $ term, $D_{\bs k}$ is a four-fermion coupling but with two key distinctions.  First, the presence of Kronecker delta function strongly constrains the scattering phase space and is attributed to integrability of the HK interaction. Second, for a quasiparticle with energy $\xi_{\bs k}$ above $\epsilon_F$ and the existence of a well defined local moment, we require the local moment $U$ to be large enough so that we have $\epsilon_d+ U> \xi_{\bs k}$. This, in addition to the fact that  $\xi_{\bs k} - \epsilon_d>0$, implies that $J(\bs k, \bs k')>0$ (antiferromagnetic) while $D_{\bs k}<0$ (ferromagnetic). Finally, the term proportional to $E_{\bs k}$ is a novel six-fermion term that is formed by the coupling of electron spins and density. The sign of $E_{\bs k}$ can be negative for values of $U, \tilde{U}$ such that $U+\epsilon_d > \tilde{U} + \xi_{\bs k}$ and $U+ 2\epsilon_d < \tilde{U} + 2\xi_{\bs k}$. This condition is satisfied for intermediate values of $U,\tilde{U}>0$. While methods employed in Refs.~\cite{Andrei1980, Andrei-RMP1983, Vigman1980, Tsvelick1983} can, in principle, be generalized to seek exact solutions of $H_K$ in Eq.~\ref{EffectiveKondo},  we leave this for future work and focus rest of the paper on performing a perturbative $T$-matrix expansion using the fully polarized HK ground state and derive scaling equations and RG flow.  \par
\begin{figure}[h!]
\includegraphics[width=3in,height=2in]{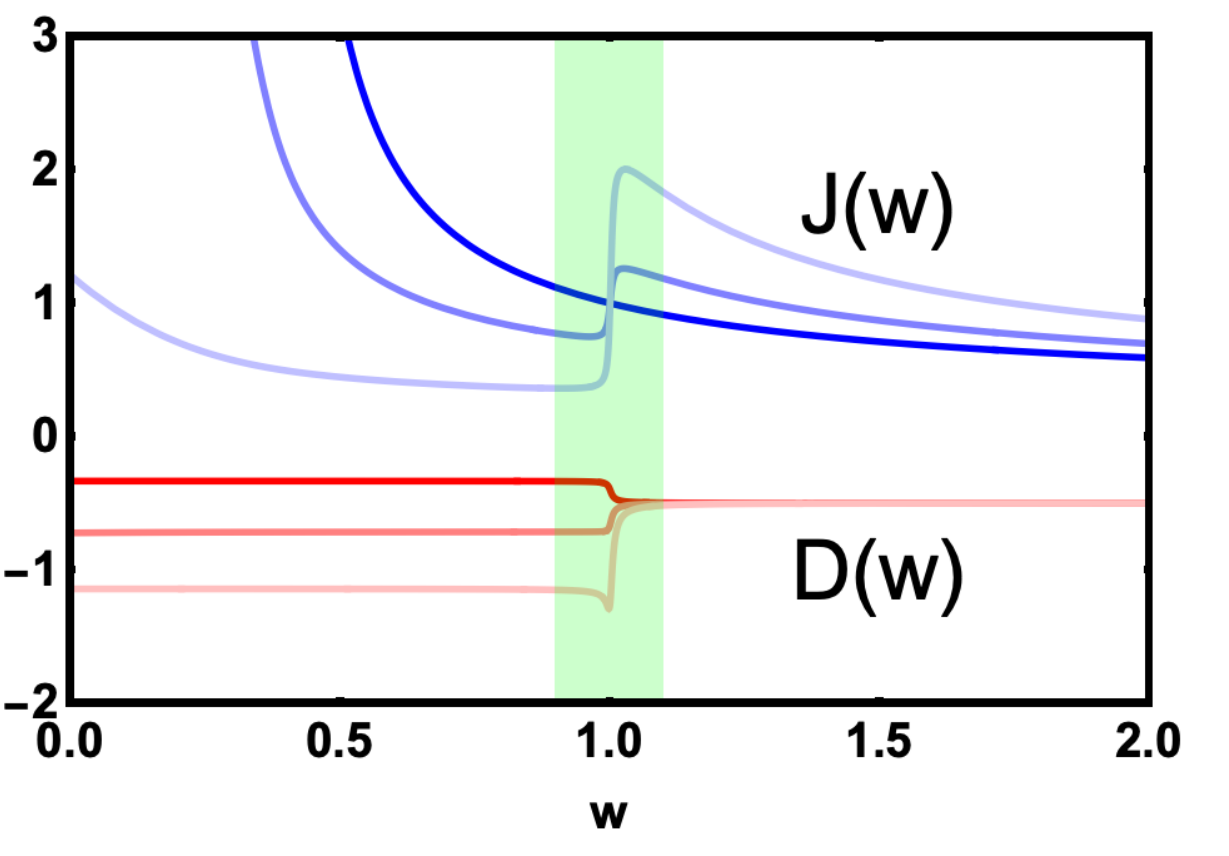} \hfill
\caption{Scaling RG flow of Kondo couplings  $J(w)$ and $D(w)$. The lighter shades correspond to higher couplings between $J(w)$ and $D(w)$.  The shaded region corresponds to resonance energy of quasiparticle denoted by $\xi_0$ in the text.}\label{RG}
\end{figure}

\textit{Scaling dimensions:} Before we set out to calculate the scattering matrix elements, it is instructive to evaluate scaling dimensions of the various terms appearing in Eq.~\ref{EffectiveKondo}.  This requires determining the scaling dimension of $c_{\bs k \sigma}$ and $d_{\sigma}$ operators which can be done in two ways: setting the scaling dimensions of either the quadratic $\xi_{\bs k}$ or the interacting $\tilde{U}$ term to be zero. Written as an action, $H_K$ has a Matsubara time variable $\tau$ that scales as $\tau \rightarrow s^{-1} \tau$ whereas the momenta scale as $ k_{\parallel} \rightarrow s k_{\parallel}$, $\bs k_{\perp} \rightarrow  \bs k_{\perp}$ for $s\rightarrow 0$. Here $k_{\parallel}, \bs k_{\perp}$ are components of momentum parallel and perpendicular to normal vector of the zero energy surface (Fermi or Luttinger surface). Fixing then the non-interacting term to be dimensionless yields $c_{\bs k} \rightarrow \frac{1}{\sqrt{s}} c_{\bs k}$ and $d_{\sigma} \rightarrow  d_{\sigma}$. From this scheme we conclude that while the $J(\bs k,\bs k')$ term is marginal (scales as $s^0$), $D_{\bs k}$ and $E_{\bs k}$ terms are relevant ($s^{-1}$) as $s$ is taken to zero. 
 If on the other hand we fix the HK interaction term $\tilde{U}$ to be dimensionless, the $c_{\bs k \sigma}$ operator also scales as $s^0$. In this scheme, the $J(\bs k, \bs k')$ and $E_{\bs k}$ terms are irrelevant ($s^{1}$) while the $D_{\bs k}$ term is marginal ($s^0$) as $s$ is taken to zero. From both these schemes, it is evident that the following poor man's scaling analysis will be controlled by the four-fermion $D_{\bs k}$ term regardless of whether the perturbation is considered about the gaussian or the strongly coupled fixed point.  \\ \newline
\textit{Scattering rate:} We now rewrite Eq.~\ref{EffectiveKondo} as $H_K = H_{HK} + H_e$ where $H_e$ consists of all the exchange terms arising from the coupling of local moments and the correlated host. For simplicity, we will also henceforth treat the couplings to be independent of momentum, $J(\bs k, \bs k') \equiv J$, $D_{\bs k} \equiv D$, $E_{\bs k} \equiv E$ where the bare quasiparticle energy is set close to $\epsilon_F$. We will also ignore the effect of the coupling $E$ in the current study. This can happen if we choose $\tilde{U}$ to be small and made to satisfy the condition $\tilde{U} \sim U - 2 \epsilon_F + 2 \epsilon_d$ with $D\neq 0$. In addition, an evaluation of the six-fermion diagrams (see Fig.~\ref{HKKondoFeyn}) shows that $E$ is effectively decoupled from $J$ and $D$ channels at low temperatures.  
 To evaluate the scattering rate, we write the Lippmann-Schwinger expansion of the $T$-matrix given by
\beq
\hat{T} = H_e + H_e \frac{1}{E_x-H_{HK} + i \eta} H_e + ..,
\eeq
where $E_x$ is the excitation energy.  To further evaluate the matrix elements  $T_{\bs k, \bs k'} = \langle \bs k \sigma|T |\bs k' \sigma' \rangle$  between \textit{in} state $|\bs k' \sigma' \rangle$ and \textit{out} state $|\bs k \sigma \rangle$, we define a `holon' excitation  $\mid \bs k' \sigma' \rangle = c_{\bs k' \sigma'}^{\dagger} (1- n_{\bs k' \bar{\sigma}'}) |\Omega\rangle$ at an energy $\xi_{0}$ close to $\epsilon_F$, where $|\Omega \rangle$ is the ground state of the HK model. As previously mentioned, the HK has a huge ground state degeneracy due to all possible spin configurations in the single occupied regions of the Brillouin zone~\cite{HK1992}. For the purposes of calculations, however, we choose the `pure' state that is maximally polarized with spin $\sigma$ as the ground state given by $|\Omega\rangle = \prod_{k\in \Omega_2}c_{\bs k \uparrow}^{\dagger}c_{\bs k \downarrow}^{\dagger}\prod_{k\in \Omega_1} c_{\bs k \sigma}^{\dagger} |0\rangle $. 
In addition, to calculate the matrix elements we note that ground state is specified by the density operator and can hence be described by the Fock basis $|\{n_j\}\rangle$. The matrix elements can therefore be factorized~\cite{Phillips2020, Danielewicz1984, Setty2018} in terms of an occupation average in the ground state 
\beq
\langle n_{\bs k \sigma}\rangle = \frac{e^{-\beta \xi_{\bs k}} + e^{-\beta (2 \xi_{\bs k}+\tilde{U})}}{1 + 2 e^{-\beta \xi_{\bs k}} + e^{-\beta (2 \xi_{\bs k}+\tilde{U})}} 
\eeq
which at zero temperature equals $\frac{1}{2}$ for $-\tilde{U}<\xi_{\bs k} <0$ and 0 (1) for $\xi_{\bs k} >0$ $(\xi_{\bs k}< -\tilde{U})$. To start, we keep only $D\neq 0$ and set all other couplings to zero. Calculating the leading order $T^{(1)}_{\bs k \bs k'}$ and next-to-leading order $T^{(2)}_{\bs k \bs k'}$ contributions to the $T$-matrix at low temperatures yields the diagonal `elastic' scattering matrix
\beq
T^{(1)}_{\bs k \bs k'}+T^{(2)}_{\bs k \bs k'}  &=& D\left( \bs S_d\cdot \vec{\sigma} \right) \left( 1- \frac{2 D}{E_x-\bar{E}_0} \right)\delta_{\bs k \bs k'} \label{DPerturbation}
\eeq
 where $\bar{E}_0$ is the renormalized ground state energy obtained by successive actions of $H_{HK}$ on $|\Omega\rangle$. To obtain the scaling equation we integrate only within an energy shell $[-w, -w+\delta w ] \cup [w-\delta w, w ] $. The second order correction to the coupling $D$ is given by
 \beq
\delta D &=& \frac{- D^2 \delta(w-\xi_0)}{E_x-\bar{E}_0} \delta w.
 \eeq
The Dirac delta function signifies the resonance condition at the energy of the excitation. This occurs entirely due to the severely restricted phase space for scattering in the $D$ term in Eq.~\ref{EffectiveKondo}. As a consequence, the \textit{in} and \textit{out} states have the same $\bs k$ (diagonal $T$-matrix) thereby constraining the internal loop momenta and energy as well.   Integrating $\delta D $ from an IR ($w'$) to UV ($w$) scale gives the running coupling
\beq
D(w') = \frac{D(w)}{1- D(w)\Theta(\xi_0 - w')(E_x-\bar{E}_0)^{-1}},
\label{DESolution}
\eeq
where $\Theta(x)$ is the Heaviside function. Thus, $D(w')$ behaves as a step-like function with $\xi_0$ marking the point of discontinuity and does not vanish at the lowest energies despite being ferromagnetic. Instead the system simply slides between two fixed points at $w =\xi_0$ (see red curve in Fig.~\ref{RG}).  The Kondo temperature when only $D\neq 0$ ($T_{KD}$) can therefore be determined by a diverging $D(w')$ which, for $w'<\xi_0$, gives $E_x-\bar{E}_0 \equiv T_{KD} \sim D$.  The modified $\beta$ function for only $D\neq 0$ is therefore given as
$$
\frac{dD}{w~d \ln w} = 
\begin{cases}
0, ~~~~~~~~~~~w\neq \xi_{0}\\
-\frac{N_0 D^2}{T_{KD}},~~~~w = \xi_{0}.
\end{cases}
$$  \\ \newline
 Turning on $J$ leads to a term that resembles Kondo effect in a Fermi liquid as well as mixing between the two couplings. The set of coupled differential equations for $J(w)$ and $D(w)$ are given by 
 \beq
J' &=& \frac{-N_0 J^2}{w} - \frac{-2 J D~\delta(w- \xi_0)}{T_{KD}}\\ 
D' &=& \frac{- D^2 \delta(w-\xi_0)}{T_{KD}} - \frac{-2 D J~\delta(w- \xi_0)}{T_{KD}}
\eeq
where $J'$ and $D'$ are derivatives of the couplings with respect to $w$. The solutions to the above set of differential equations is shown in Fig.~\ref{RG} with initial conditions $J(w)>0$ and $D(w) < 0$ in the UV limit. In the absence of any mixing between the two (dark blue and red curves in Fig.~\ref{RG}), $J(w)$ exhibits the usual Kondo log divergence and $D(w)$ shows a step-like flow as indicated above. With an increase in mixing between the couplings (lighter curves in Fig.~\ref{RG}), $J(w)$ also inherits a step-like RG flow at $w=\xi_0$ and the $w=0$ divergence is cut-off to a finite value. Therefore, unlike the Fermi liquid Kondo case where perturbation theory breaks down at lowest energies due a divergence in $J(w)$, a well-defined expansion is still valid in the Kondo-HK model.  \par
\textit{Discussions:}
As alluded to above, integrability of the HK model is at the heart of the step-like RG flow pattern shown in Fig.~\ref{RG}. In the canonical Kondo effect, the internal loop quantum numbers label virtual states and are hence arbitrary. In the Kondo-HK model, however, the Kronecker delta functions appearing in Eq.~\ref{EffectiveKondo} strongly constrain these quantum numbers to values set by the \textit{in} and \textit{out} states with no role for virtual states; therefore, when only $D\neq 0$, the scattering matrix is diagonal in momentum. This is a direct consequence of extensively many conserved density operators, one for each $\bs k $ and a characteristic feature of integrability~\cite{Doyon2008}. The step-like discontinuity in the flow occurs when the scaling energy equals the energy of the quasiparticle that undergoes scattering while the couplings determine the ground state and excitation gap at that energy.  Thus properties of the system drastically differ if they are probed at an energy slightly above (UV) or below (IR) this scale, i.e., if the scattered particle is effectively massless or massive with respect to the probe energy. \par
An analogy can be drawn between the physics described above and  diagonal resonance scattering in relativistic integrable quantum field theories put forward by Zamolodchikov and others~\cite{Zamolodchikov1991, Zamolodchikov1991-2, Zamolodchikov1991-3, Zamolodchikov1991-4, Zamolodchikov1991-5, Martins1992, Ravanini1993, Castro2000, Leclair2003, Castro2004, Takacs2015, Pimenta2017, Lee2020}. An example is the homogenous Sine-Gordon model~\cite{Miramontes1998, Miramontes2000} (for a gentler introduction to these topics, the reader is encouraged to follow Refs.~\cite{Miramontes2002, SGModel2014}) which is an integrable deformation of a conformal field theory (CFT). The model describes scattering of solitonic solutions of the Sine-Gordon model labelled by certain quantum numbers (mass, resonance parameters etc). Like earlier, integrability of the model implies dynamics are constrained resulting in a diagonal ``elastic" scattering matrix. The ground state energy is determined by the \textit{effective central charge} $c(w)$ -- a quantity that flows with the RG scale and is evaluated from solutions of the Thermodynamic Bethe Ansatz (TBA). The effective central charge  hence plays the analogous role of HK-Kondo couplings. Crucially, there is a step-like jump in $c(w)$ when the RG scale equals the particle mass determined by the \textit{in} and \textit{out} solitonic states.  More generally, multiple plateaux are possible (``staircase" models) when several particles can scatter to give rise to well separated mass scales; in this scenario, the system slides from one CFT to the next depending on the probe energy and eventually settles into an IR fixed point.  It will be interesting to reproduce such staircase patterns in the Kondo-HK model when multi-particle scattering is taken into account and will be the subject of future work. \par
\bibliography{HK-Kondo.bib} 
\newpage
\onecolumngrid
\section{Appendix }
\begin{figure}[h!]
\includegraphics[width=3in,height=3in]{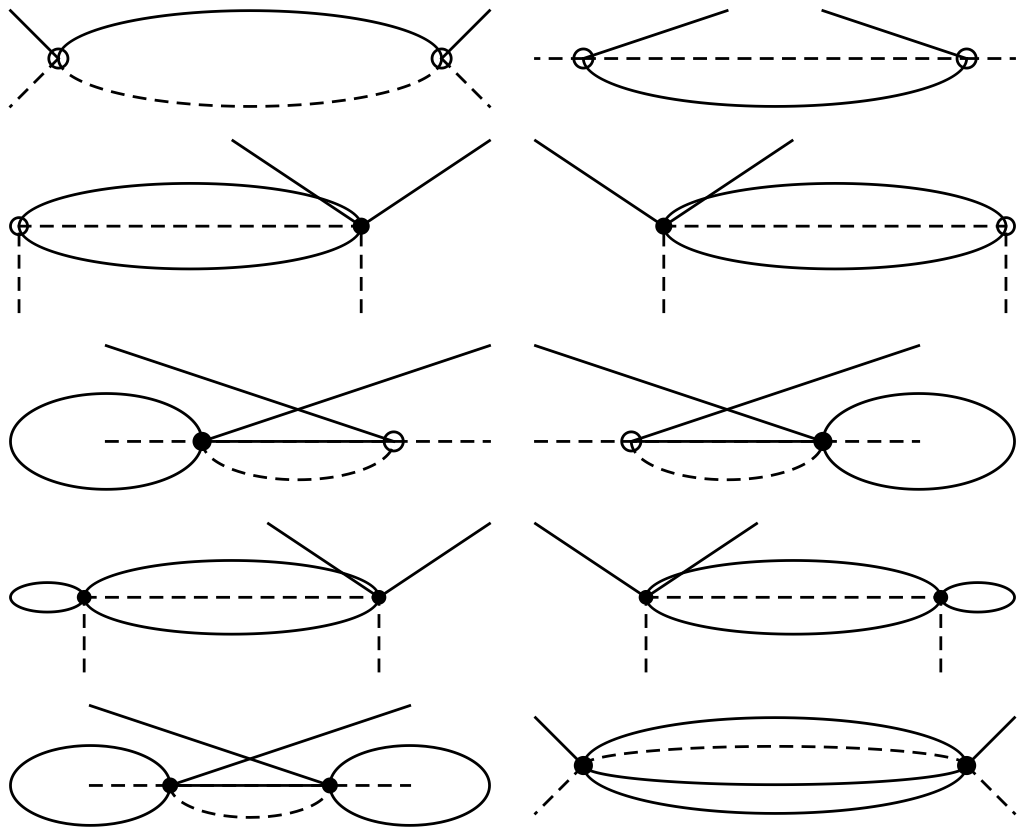} \hfill
\caption{Feynman diagrams contributing to scattering probabilities between interacting conduction electrons (solid lines) and impurity spins (dashed lines). The four (six) fermion vertices are denoted by open circles (solid disks).  }\label{HKKondoFeyn}
\end{figure}
In the Appendix, we expand on details of calculations appearing in the main text. Fig.~\ref{HKKondoFeyn} shows a summary of distinct Feynman diagrams contributing to the scattering amplitude at second order in the perturbation. The top row contains only four-fermion vertices (open circles) like those that appear in the canonical Kondo effect for a Fermi liquid. These diagrams contribute to $D(w)^2$, $J(w)^2$ or $D(w) J(w)$ terms in the RG equations. The second and third rows are mixed diagrams that contain both six-fermion (solid disks) and four-fermion vertices. These diagrams contribute to mixing terms of the form $D(w) E(w)$ or $J(w) E(w)$. Finally, the bottom two rows contain purely six-fermion vertices and contribute to $E(w)^2$ terms in the $T$-matrix.
To understand the step-like RG flow of $D(w)$ outlined in the main text, we recall the original perturbative argument for magnetic moments in a Fermi liquid. The leading order contribution to the impurity-conduction electron coupling is
\beq
T^{(1)}_{\bs k \bs k'} = J(\bs S_d.\vec{\sigma}) (1- f_{\bs k}) (1- f_{\bs k}').
\eeq
To second order we have
\beq
T^{(2)}_{\bs k \bs k'} = J^2 \sum_{\alpha} \sigma_{\sigma \alpha}^i\sigma_{\alpha \sigma'}^j S_d^i S_d^j \sum_{\bs p} \frac{(1- f_{\bs k})(1- f_{\bs p}) (1- f_{\bs k'})}{E_x - E_0 - \xi_{\bs{p}}} -J^2 \sum_{\alpha}\sigma_{\alpha \sigma'}^i \sigma_{\sigma \alpha}^j S_d^i S_d^j \sum_{\bs p} \frac{(1- f_{\bs k})f_{\bs p} (1- f_{\bs k'})}{E_x - E_0 - \xi_{\bs{k}} + \xi_{\bs{p}} - \xi_{\bs{k}'} }
\eeq
where $E_0$ is the ground state energy. We next break up $\bs p$ integral into states below and above the Fermi energy and perform the relevant integrals.  Utilizing the relations $[\sigma^i, \sigma^j ] =2 i \epsilon_{ijk} \sigma^k$ and $2 i J^2 S_d^i S_d^j \epsilon_{ijk} \sigma^k = -4 J^2 (\bs S_d.\vec{\sigma})$ where $\epsilon_{ijk}$ are the Levi-Civita constants, we can combine the two terms proportional to $\bs S_d.\vec{\sigma}$. At low temperatures and $\bs k, \bs k'$ states above the fermi energy we can use $f_{\bs k}, f_{\bs k}' \sim 0$ and write 
\beq
T^{(1)}_{\bs k \bs k'} + T^{(2)}_{\bs k \bs k'} =  J (\bs S_d.\vec{\sigma}) \left( 1+ 2 N_0 J \ln \left |\frac{w}{E_x - E_0} \right| \right).
\eeq
The perturbative correction to the original coupling $J$ is hence proportional to $J^2$, i.e., $\delta J= -2J^2 N_0 d \ln w$. We now apply a similar argument for the coupling $D$. 
For the $T$-matrix contributions from the $D$ terms, we have additional Kronecker functions appearing in the second line of Eq.~\ref{EffectiveKondo}. These constrain the incoming, outgoing and internal loop momenta as $\bs p = \bs k = \bs k'$ making the $\bs p$ integral trivial. Therefore, the first and second order terms are  
\beq
T^{(1)}_{\bs k \bs k'} &=& D (\bs S_d.\vec{\sigma})  (1- f_{\bs k})^2 \\
T^{(2)}_{\bs k \bs k'} &=& \delta_{\bs k \bs k'}D^2 \sum_{\alpha} \sigma_{\sigma \alpha}^i\sigma_{\alpha \sigma'}^j S_d^i S_d^j  \frac{(1- f_{\bs k})^3}{E_x - \bar{E}_0} -\delta_{\bs k \bs k'} D^2 \sum_{\alpha}\sigma_{\alpha \sigma'}^i \sigma_{\sigma \alpha}^j S_d^i S_d^j \frac{(1- f_{\bs k})^2 f_{\bs k}}{E_x - \bar{E}_0}. \label{SecondOrderDTerm}
\eeq
For an excitation with energy $\xi_{\bs k} \gtrsim 0$, we have $f_{\bs k}\sim 0$ at low temperatures and only the first term in $T^{(2)}_{\bs k \bs k'} $ above contributes. Using the relation $\sigma^i \sigma^j = \delta_{ij} + i \epsilon_{ijl} \sigma_l$ for the $\sigma$ operator (similarly for the $\bs S_d$ operator) and keeping only the terms proportional to $(\bs S_d.\vec{\sigma}) $, we have Eq.~\ref{DPerturbation} of the main text
\beq
T^{(1)}_{\bs k \bs k'} + T^{(2)}_{\bs k \bs k'} = D (\bs S_d.\vec{\sigma}) \left(1- 2 \frac{D}{E_x - \bar{E}_0}\right)\delta_{\bs k \bs k'}. 
\eeq
To derive the scaling equations, we rewrite Eq.~\ref{SecondOrderDTerm} by explicitly inserting a Dirac delta function and converting the momentum sum into an energy integration over a narrow energy window $[-w, -w+\delta w]\cup [w-\delta w, w] $. Since only the first term contributes, we have
\beq
T^{(2)}_{\bs k \bs k'} =\delta_{\bs k \bs k'}D^2\int_{w-\delta w}^w d \xi_{\bs p} \sum_{\alpha} \sigma_{\sigma \alpha}^i\sigma_{\alpha \sigma'}^j S_d^i S_d^j  \frac{(1- f_{\bs k})^2 (1- f_{\bs p}) }{E_x - \bar{E}_0} \delta(\xi_{\bs p} - \xi_{\bs k}).
\eeq
The integral above is non-zero only when $w-\delta w < \xi_{\bs p} = \xi_{\bs k}\equiv \xi_0 < w$. Using spin identities like before and collecting terms proportional to $(\bs S_d.\vec{\sigma}) $ and we are left with
 \beq
 T^{(2)}_{\bs k \bs k'} =-\delta_{\bs k \bs k'} D^2 \frac{(\bs S_d.\vec{\sigma})}{E_x - \bar{E}_0} \Theta(w-\xi_0) \Theta(\xi_0 - w +\delta w) \simeq -\delta_{\bs k \bs k'} D^2 \frac{(\bs S_d.\vec{\sigma})}{E_x - \bar{E}_0} \delta(\xi_0 - w)\delta w.
 \eeq
As described in the main text, this implies the correction to the original coupling $D$ due to the perturbation can be expressed as a differential equation using a Kronecker function given by
$$
\frac{dD}{w~d \ln w} = 
\begin{cases}
0, ~~~~~~~~~~~w\neq \xi_{0}\\
-\frac{N_0 D^2}{E_x - E_0},~~~~w = \xi_{0}.
\end{cases}
$$
The solution to the differential equation above can be obtained readily by integrating $D(w)$ from a low energy scale $w'$ to a high energy scale $w'$ resulting in Eq.~\ref{DESolution} of the main text.  A similar argument follows for the mixed terms proportional to $J(w) D(w)$ when $J(w)$ is turned on. The perturbative correction to $D(w)$ $\left(J(w)\right)$ due to $J(w)$ $\left(D(w)\right)$ looks like Eq.~\ref{SecondOrderDTerm} where $D(w)^2$ is simply replaced by the product $J(w) D(w)$. 
\end{document}